\documentclass[a4paper,10pt]{article}
\usepackage{bm}
\usepackage{booktabs}
\usepackage{array}
\usepackage{latexsym}  
\usepackage{dcolumn}
\usepackage{amsmath,amsfonts,amssymb}
\usepackage{graphicx,epsfig}
\usepackage{color}
\usepackage{verbatim}
\usepackage{psfrag}

\def\app#1{{Appendix~\ref{#1}}}
\def\sec#1{{Section~\ref{#1}}}
\def\eq#1{{Eq.~(\ref{#1})}}

\def\Cal{\mathcal}

\title{Complex Effective Path: A Semi-Classical Probe of Quantum Effects}

\author{Suprit Singh\footnote{suprit@iucaa.ernet.in} ~ and T.~Padmanabhan\footnote{nabhan@iucaa.ernet.in}\\
IUCAA, Pune University Campus, Ganeshkhind,\\
 Pune 411007, INDIA. \\ 
} 

\date{\today}

\begin{document}

\maketitle

\begin{abstract}
We discuss the notion of an effective, average, quantum mechanical path which is a solution of the dynamical equations obtained by extremizing the quantum effective action. Since the effective action can, in general, be complex, the effective path will also, in general, be complex. The imaginary part of the effective action is known to be related to the probability of particle creation by an external source and hence we expect the imaginary part of the effective path also to contain information about particle creation. We try to identify such features using simple examples including that of effective path through the black hole horizon leading to thermal radiation. Implications of this approach are discussed. 
\end{abstract}


\section{Introduction} \label{sec:intro}

The study of a quantum mechanical system interacting with an externally specified classical background is of importance in several physical contexts. Such an external classical source will, in general, lead to vacuum polarization and particle production. Well known examples of these phenomena occur in the study of Schwinger effect \cite{schwinger,effaction,tpaspects}, particle creation in expanding universe \cite{effaction,pexpuniv} and black hole evaporation \cite{effaction,bhevap}.
A powerful technique to study such external source problems is that of the effective action which captures the quantum effects through a c-numbered effective action functional, $S_{\rm eff}\equiv\Gamma$ of the dynamical variables \cite{effaction,tpaspects}. In general, the effective action will be a complex quantity with its real and imaginary parts being related to vacuum polarization and  particle production respectively. Conventionally, one writes down the effective dynamical equations for the system by varying \textit{only the real part} of the effective action thereby identifying the quantum  corrections to the classical equations. For example, in the case of electromagnetic field, such an approach will lead to the Euler-Heisenberg effective action which can provide quantum corrections to classical Maxwell's equations \cite{tpaspects,euhe}.
The imaginary part of the effective action is not usually considered in such a variational principle since in many applications the effect of vacuum polarization dominates over that due to particle production.

It is interesting to ask whether one can extend the above formalism to include the effects of imaginary part of the effective action as well since it could, potentially, provide a formal procedure for handling the back reaction due to particle production. The obvious procedure would be to look for the solutions of $\delta \Gamma=0$ where both the real and imaginary part of $\Gamma$
are retained. These equations will, in general, be complex rendering  the solutions also to be complex.
For example, in the elementary context of non-relativistic quantum mechanics, such a solution is the effective average path $X(t;x_2,t_2; x_1,t_1)$ obeying the appropriate  boundary conditions  at the end points. This function will, in general, be complex and one would presume that its imaginary part will contain some information about the particle production due to the external source. The primary aim of this paper is to investigate the properties of this function.

It might seem that, since the effective path $X(t)$ is a solution to the effective field equation $\delta \Gamma=0$, it can be determined only after $\Gamma$ is explicitly obtained which in turn would depend on the system under consideration. We shall see, however, that there is a simple way of characterizing $X(t)$ as a path integral average of all paths so that it can be expressed as an integral involving the standard path integral kernel. (This idea was first developed in \cite{brown} but we could not find any follow up of this idea in the literature, hence we shall provide fair amount of details of the approach in this paper.) This is the approach we shall use to investigate the properties of $X(t)$ in this paper.

In the above discussion, we have made a correspondence between the imaginary part of the
action with the existence of phenomena like particle production or vacuum
instability. This is indeed the case for the specific examples which we
will be concerned with in this paper. However, it should be mentioned that
one can have situations in which imaginary terms  arise in the
Euclidean action due to other reasons, which are usually topological. One
key example of this is in the context of terms in the Minkowski action
which  are odd under time reversal.  When analytically continued to the
Euclidean sector such terms can give rise to an imaginary part in the 
Euclidean action. Examples of this include topological terms, Wess-Zumino
term, Chern-Simons term  etc. (see for e.g. ref. \cite{Alexanian}). We
will not be concerned with actions containing such terms in this paper.

The plan of the paper is as follows. In~\sec{seceffactnandpath} we briefly review the concept of the effective path as a solution to the effective action equations and its connection with the path integral. We evaluate the effective path in the case of  a harmonic oscillator interacting  with an external source in~\sec{secho} We show that the effective path for this case is complex and its modulus square can be related to the total energy input into the system by the external source due to the production of particles. (Interestingly, the effective path in this case is similar to a complex quantity constructed by Landau and Lifshitz in \cite{ll1} to solve the problem of a forced oscillator in \textit{classical} mechanics.)  We next consider (\sec{secinvsqr}) the effective path for a non-quadratic system with potential $-1/x^2$ and evaluate the modulus square in a suitable approximation. We find that this quantity has a rather curious form in that it contains a `Planck spectrum'. We know, however, from previous work \cite{TPSreeni} that  the  problem of thermal radiation from a horizon can be mapped to that of quantum mechanics in an inverse square potential. We study (\sec{sechorizon}) the properties of the effective path in this context and show that its modulus square can be related to the Hawking temperature (except for a factor of 2, the origin of which has been discussed extensively in the literature \cite{factortwo}). In~\app{appeffpathclassofinvsqr} we also extend the results of \cite{brown} to a more general class singular potentials with the hope that it will be of future use.

\section{Effective action and the concept of effective path}
\label{seceffactnandpath}
We shall begin by introducing the notion of effective path and its relation to the standard effective action. We shall work in the context of point quantum mechanics because it is adequate for our purposes; the generalization to a field theoretic context is conceptually similar though mathematically more involved. In the context of point quantum mechanics, the path integral kernel describing the system is given by the Feynman path integral
\begin{align}
K(x_2,t_2|x_1,t_1) &= \langle x_2,t_2|x_1,t_1\rangle \nonumber\\
&= \int \mathcal{D}x(t)\,\exp\frac{i}{\hbar}S[x(t)]  
\end{align}
where the sum is over all the paths satisfying the indicated boundary conditions. This suggests a very natural definition of an effective average path using the path integral average:
\begin{equation}
 X(t) \equiv \frac{\int \mathcal{D}x\,x\,\exp[iS/\hbar]}{\int \mathcal{D}x\,\exp[iS/\hbar]}
  = \frac{\langle x_2,t_2|\hat{x}(t)|x_1,t_1\rangle}{\langle x_2,t_2|x_1,t_1\rangle}.
\label{piav}
\end{equation} 
In terms of the path integral kernel, the effective path can be expressed as 
\begin{align}
\label{pthkern}
X(t) &=  \frac{\langle x_2,t_2|\hat{x}(t)|x_1,t_1\rangle}{\langle x_2,t_2|x_1,t_1\rangle} \nonumber\\ 
&= \frac{\int_{-\infty}^{\infty} \mathrm{d}x\,x\,K(x_2,t_2|x,t)\,K(x,t|x_1,t_1)}{K(x_2,t_2|x_1,t_1)}.
\end{align}

We  can evaluate this function once we know the path integral kernel for the system. While the path integral average in \eq{piav} appears to be a natural quantity to define, it should be noted that --- being a transition matrix element rather than the expectation value of an operator --- it is in general a complex quantity (which is probably why it has not received any attention in the literature; we could not find any published study of this quantity except in ref. \cite{brown}). But what makes $X(t)$ important is that it is a solution to the effective action equations $\delta \Gamma=0$ including the imaginary part of the effective action. We shall now provide a short proof of this claim for the sake of completeness.

The standard procedure for defining the effective action is as follows. We introduce an external source $J(t)$ and define 
\begin{align}
\label{genfnctnl}
\exp\frac{i}{\hbar}W[J(t)] &= \langle x_2,t_2|x_1,t_1\rangle_J\nonumber\\ 
&= \int \mathcal{D}x(t)\,\exp\frac{i}{\hbar}\left(S[x(t)] + \int dt J(t) x(t)\right)  
\end{align}
where $W[J]$ is the generating functional for Green functions. Functional differentiation of the generating function with respect to $J$ then immediately leads to something very similar to the quantity in \eq{piav}, and, of course, is used in the literature: 
\begin{equation}
\label{3}
X[J] \equiv \frac{\delta W[J]}{\delta J} = \frac{\langle x_2,t_2|\hat{x}(t)|x_1,t_1\rangle_J}{\langle x_2,t_2|x_1,t_1\rangle_J}
\end{equation}
which is the effective average path of the system for the specified boundary conditions but in the presence of the external source. This relation can be inverted to get $J =J[X]$ and hence allows us to naturally define a functional of $X$, $\Gamma[X]$, as the Legendre transform of $W[J]$ with respect to $J$ as 
\begin{equation}
\Gamma[X] \equiv W[J] - \int J(t) X(t) dt
\end{equation}
where $J$ is now considered a functional of $X$. It is easy to see that the functional derivative of $\Gamma[X]$ is given by
\begin{equation}
\frac{\delta \Gamma[X]}{\delta X} =  - J.
\end{equation}
Thus, the extremum condition for effective action, giving the effective, quantum corrected, dynamical equation, $\delta_X \Gamma = 0$, implies $J=0$. Therefore its solution  is just $X[J]$ evaluated at $J=0$ which is
\begin{equation}
X[0] \equiv \left. \frac{\delta W[J]}{\delta J} \right\arrowvert_{J=0} = \frac{\langle x_2,t_2|\hat{x}(t)|x_1,t_1\rangle}{\langle x_2,t_2|x_1,t_1\rangle},
\end{equation}
the effective path given by \eq{piav} in the absence of the source. Since the effective action can, in general, be complex, it follows that the complex nature of $X(t)$ contains information about the complex nature of effective action. It is this aspect of the effective path which we will focus our attention on using simple examples.

\section{Effective path for forced harmonic oscillator} 
\label{secho}

We begin by considering the case of a harmonic oscillator coupled linearly to an external source, $J(t)$. We will assume that $J(t)$ was switched on and switched off sufficiently fast when $t\to \pm \infty$.
The oscillator evolves from the initial vacuum state in the asymptotic past  to the final  vacuum state in the asymptotic future. The in-out vacuum-to-vacuum amplitude can be calculated \cite{mukhanov} to be 
\begin{equation}
\label{7}
\langle 0_{\mathrm{out}}|0_{\mathrm{in}}\rangle = \exp \left(-\frac{1}{4\hbar\omega}|\tilde{J}(\omega)|^2 \right)
\end{equation}
where $\tilde{J}(\omega)$ is the Fourier mode of $J(t)$ at the oscillator's natural frequency, $\omega$. Since the oscillator can only absorb quanta at its natural frequency $\omega$, we see that only the fourier mode of $J(t)$ at the natural frequency of the oscillator is relevant for particle production.

The calculation of the effective action for this system proceeds in a straightforward manner. By definition,
\begin{equation}
\exp[iW[J(t)]/\hbar] = \int \Cal{D}x(t) \exp i S[J(t),x(t)]/\hbar. 
\end{equation}
where the action is given by 
\begin{equation}
S[x,J] =  -\int \left(\frac{1}{2} x\hat{D}x - J(t)\, x\right)dt
\end{equation}
with $\hat{D}$ as the standard harmonic oscillator differential operator. The path integral for the system can be computed by elementary procedures to give
\begin{equation}
\exp[iW[J(t)]/\hbar] = (\det D)^{-\frac{1}{2}} \exp\frac{i}{2\hbar}\int\mathrm{d}t\int\mathrm{d}t'\,J(t)G_F(t,t')J(t'). 
\end{equation}
where $G_F$ is the Feynman Green function for the harmonic oscillator.
The corresponding generating function is given by 
\begin{align}
W[J(t)] &= \frac{1}{2}\int J(t)J(t')G_F(t,t')\mathrm{d}t\mathrm{d}t'\nonumber\\
            &= \frac{i}{4\omega}|\tilde{J}(\omega)|^2 + \int J(t)J(t')\frac{\sin \omega|t-t'| }{4\omega}\mathrm{d}t\mathrm{d}t'     
\end{align}
apart from a $J$-independent part from the $(\det D)^{-1/2}$ which is irrelevant for our purpose. Using the definition
\begin{equation}
|\tilde{J}(\omega)|^2 = \int\mathrm{d}t\mathrm{d}t' e^{i\omega(t-t')}J(t)J(t') 
\end{equation}
we see that the imaginary part of $\Gamma[J]$ is precisely the in-out matrix element (which, of course, can be evaluated directly in this simple case):
\begin{equation}
\langle 0_{\mathrm{out}}|0_{\mathrm{in}}\rangle = \exp \left(-\frac{1}{4\hbar\omega}|\tilde{J}(\omega)|^2 \right). 
\label{inout}
\end{equation}
The transition probability is the modulus squared of the amplitude
\begin{equation}
|\langle 0_{\mathrm{out}}|0_{\mathrm{in}}\rangle|^2 = \exp \left(-\frac{1}{\hbar\omega}\frac{|\tilde{J}(\omega)|^2}{2}\right)  
\end{equation}
from which we can  read off the energy transferred to the oscillator by the external source to be
\begin{equation}
 \mathcal{E} =  \frac{1}{2}|\tilde{J}(\omega)|^2.
\end{equation}
Thus, a time-dependent source with non-zero $\tilde{J}(\omega)$ driving a harmonic oscillator does produce transitions of eigenstates so that the `in' and the `out' states are different with the amplitude given by imaginary part of the effective action.  

All this is fairly standard and we shall now introduce the effective path for the system as a solution to the effective dynamical equations obtained by extremizing the effective action. It is obvious that while the equation of motion is the same as the classical one,
\begin{equation}
\ddot{x} + \omega^2 x = J(t) 
\end{equation}
its solution should be now obtained in terms of the Feynman Green function (rather than the standard retarded Green function) which makes the effective path complex:
\begin{equation}
X(t) = \int \mathrm{d}t' G_F(t,t')J(t') + x^H(t).
\end{equation}
Here $x^H(t)$ is the solution to the homogenous equation of motion without the external source. 
The oscillator in the absence of external force evolves as  
\begin{equation}
x^H_{cl}(t) = x_1 \frac{\sin\omega(t_2-t)}{\sin\omega(t_2-t_1)} + x_2\frac{\sin\omega(t-t_1)}{\sin\omega(t_2-t_1)}  
\end{equation}
between the boundary points $x _1(t_1)$ and $x_2(t_2)$. Letting $t_2=-t_1=T$ and taking the limit $iT\rightarrow\infty$, we see that
 $x^H_{cl}$ vanishes in our case when we consider sufficiently large time intervals.
This gives the effective path to be 
\begin{align}
X(t) &= \int\mathrm{d}t' G_F(t,t')J(t') = \int \mathrm{d}t' J(t')\frac{e^{-i\omega |t-t'|}}{2\omega}\nonumber\\
     &= \int\mathrm{d}t' J(t')\frac{i}{2\omega}\left[e^{-i\omega(t-t')}\theta(t-t') + e^{i\omega(t-t')}\theta(t'-t)\right]\nonumber\\
     &= \int_{-\infty}^{t}\mathrm{d}t' J(t')\frac{i}{2\omega}e^{-i\omega(t-t')} + \int_{t}^{\infty}\mathrm{d}t' J(t')\frac{i}{2\omega} e^{i\omega(t-t')}
\end{align}
with the real and imaginary parts
\begin{align}
\mathrm{Re}X(t) &= \int_{-\infty}^{t}\mathrm{d}t' J(t')\frac{\sin \omega(t-t')}{\omega} - \frac{1}{2}\int_{-\infty}^{\infty}\mathrm{d}t' J(t')\frac{\sin \omega(t-t')}{\omega}\nonumber\\
&= x_{cl}(t) -  \frac{1}{2}\int_{-\infty}^{\infty}\mathrm{d}t' J(t')\frac{\sin \omega(t-t')}{\omega}\\  
\mathrm{Im}X(t) &= \frac{1}{2}\int_{-\infty}^{\infty}\mathrm{d}t' J(t')\frac{\cos \omega(t-t')}{\omega}
\end{align}
where $x_{cl}(t)$ is the classical solution to the driven oscillator evaluated with retarded boundary conditions.
\begin{align}
x_{cl}(t) &= \int\mathrm{d}t' G_R(t,t')J(t') = \int \mathrm{d}t' J(t')\frac{\sin\omega(t-t')}{\omega}\theta(t-t')\nonumber\\  
          &= \int_{-\infty}^{t} \mathrm{d}t' J(t')\frac{\sin\omega(t-t')}{\omega}. 
\end{align}
It is obvious that the net effect of the source is to introduce an imaginary part to $X(t)$ and modify the real part by an extra term.

Since we have already shown that the effective path $X(t)$ is a solution to the effective action equations, one can also compute the effective action for our system by evaluating it for the effective complex path given above. An elementary calculation shows that the result is given by
\begin{equation}
\Gamma[X_{\mathrm{eff}}] =  -\frac{1}{2}\int\mathrm{d}t\,\left(X_{\mathrm{eff}}J - 2JX_{\mathrm{eff}}\right) = \frac{1}{2}\int\mathrm{d}t\,JX_{\mathrm{eff}}  
\end{equation}
so that 
\begin{align}
\mathrm{Im}\,\Gamma[X_{\mathrm{eff}}] &= \frac{1}{2}\int\mathrm{d}t\, J\, \mathrm{Im}X_{\mathrm{eff}}\\
&= \int\mathrm{d}t\mathrm{d}t'\, \frac{\cos \omega(t-t')}{4\omega} \,J(t)J(t') = \frac{1}{4\omega}|\tilde{J}(\omega)|^2
\end{align}
which agrees with the result obtained in \eq{inout}.

We will now highlight the above aspects with an explicit example. Consider the source $J(t) = |t|e^{-\lambda |t|}$, which is chosen specifically to distinguish  the cases in which the particle production occurs from those in which it does not. We have seen that the energy that goes into the system from external source is proportional to the modulus square of fourier mode of the source evaluated at  $\omega$, natural frequency of the oscillator. For our choice of $J(t) = |t|e^{-\lambda |t|}$ we have:
\begin{equation}
|\tilde{J}(\omega)|^2 = \frac{(\lambda^2 - \omega^2)^2}{(\omega^2 + \lambda^2)^4} 
\end{equation}
which vanishes for the parameter $\lambda = \omega$ and hence there is no particle production in that case. We have tabulated the results for the two cases, one with a general $\lambda$ and the other with $\lambda =\omega$: 

\begin{table}[ht]
\centering
\begin{tabular}{cp{6.5cm}p{3.5cm}}
\toprule
$J(t)$ & $|t|e^{-\lambda|t|}$ & $|t|e^{-\omega|t|}$\\
\toprule
\\
$x_{cl}(t)$ & $\left(\frac{2(\lambda^2 - \omega^2)\sin\omega t} {\omega(\omega^2 + \lambda^2)^2} + \frac{e^{-\lambda t}(\lambda(2+\lambda t)+\omega^2 t)}{(\omega^2 + \lambda^2)^2}\right)\theta(t) - 
\left(\frac{e^{\lambda t}(\lambda(-2+\lambda t)+\omega^2 t)}{(\omega^2 + \lambda^2)^2}\right) \theta(-t)$ &  $\frac{e^{-\omega t}(\omega(2+\omega t)+\omega^2 t)}{4\omega^4}\theta(t) - \frac{e^{\omega t}(\omega(-2+\omega t)+\omega^2 t)}{4\omega^4}\theta(-t)$\\
\\
$\mathrm{Re}X(t)$ & $\left(\frac{(\lambda^2 - \omega^2)\sin\omega t}{\omega(\omega^2 + \lambda^2)^2} + \frac{e^{-\lambda t}(\lambda(2+\lambda t)+\omega^2 t)}{(\omega^2 + \lambda^2)^2}\right)\theta(t) -
\left(\frac{(\lambda^2 - \omega^2)\sin\omega t}{\omega(\omega^2 + \lambda^2)^2} +\frac{e^{\lambda t}(\lambda(-2+\lambda t)+\omega^2 t)}{(\omega^2 + \lambda^2)^2}\right)\theta(-t)$ & $x_{cl}|_{\lambda = \omega}$\\
\\
$\mathrm{Im}X(t) $&$ \frac{(\lambda^2 - \omega^2)\cos\omega t}{\omega(\omega^2 + \lambda^2)^2} $ & $0$\\
\\
$|X|^2_{t=\infty}$ & $\frac{(\lambda^2 - \omega^2)^2}{\omega^2(\omega^2 + \lambda^2)^4}$ & $0$\\
\\
$|\tilde{J}(\omega)|^2 = 2 \mathrm{Im}\,W$ & $\frac{(\lambda^2 - \omega^2)^2}{(\omega^2 + \lambda^2)^4}$ & $0$\\
\\\bottomrule
\end{tabular}
\end{table}

It is obvious that the imaginary part of the effective path is related to the particle production and vanishes when there is no particle production. Further, when $\lambda t\to \infty$ we can approximate the real and imaginary parts of $X(t)$ by
\begin{equation}
\label{Xreimho}
 \mathrm{Re}X(t)\approx \frac{(\lambda^2 - \omega^2)\sin\omega t}{\omega(\omega^2 + \lambda^2)^2};\quad
 \mathrm{Im}X(t)= \frac{(\lambda^2 - \omega^2)\cos\omega t}{\omega(\omega^2 + \lambda^2)^2}.
\end{equation} 
It follows that 
\begin{equation}
\mathcal{E} = \frac{\omega^2 |X|^2_{t\to \infty} }{2}= \frac{|\tilde{J}(\omega)|^2}{2}= \mathrm{Im}\,W 
\end{equation} 
giving a direct relation between the particle production rate and the squared modulus of the effective path. It is also worth mentioning here that the effective path which we get as the solution of effective action equation of motion, interestingly, gives an interpretation to the complex quantity,
\begin{equation}
\label{llconstruct}
\xi (t) = \dot{x} + i\, \omega x
\end{equation}
constructed in \cite{ll1} purely as a mathematical trick for solving the problem of forced harmonic oscillator. The energy input into the system in terms of $\xi$ is 
\begin{equation}
\mathcal{E} = \frac{|\xi(\infty)|^2}{2}.
\end{equation}
We can identify the corresponding real and imaginary parts in $X(t)$ and $\xi(t)$ apart from a factor of $\omega$. This elementary illustration shows that even in the context of such a simple system the concept of effective path can be related to a tangible result. 


\section{Inverse square potential in quantum mechanics and applications to horizon thermodynamics}

The results in the above case are rather simple because the coupling was linear. We next investigate the complex path formalism in the case of a nontrivial example, involving one-dimensional inverse square potential. The primary   motivation for this arises from the fact that the problem of a scalar field in Schwarzschild background --- and, more generally, in any spacetime in which the near horizon geometry can be approximated as Rindler --- can be reduced to dynamics of a particle in an inverse square potential across the singularity.  We explore the nature of the effective path  in this potential and show that it has some curious features which find application to the problem of  black hole evaporation. 


\subsection{Complex effective path for the inverse square potential}
\label{secinvsqr}
We will consider an inverse square potential of the form
\begin{equation}
\label{potential}
V(x) = -\frac{\hbar^2}{2m}\left(a^2 + \frac{1}{4}\right)\frac{1}{x^2} = -\frac{\tilde{\alpha}}{x^{2}}. 
\end{equation}
where $a,\tilde{\alpha}$ are constants. Since $a$ is real, $\tilde{\alpha}>\hbar^2/8m$.
To calculate the effective path in this case, we will use the path integral average. The kernel for a particle to propagate from points $(x_1,t_1)$ to $(x_2,t_2)$ in an inverse square potential, $V = -\tilde{\alpha} x^{-2}$  is given by (see~\app{appinvsqrkernel} for details),
\begin{equation}
\label{kernel}
K(t_2,x_2|t_1,x_1) = e^{-\frac{1}{2}i\pi(\gamma + 1)}\left(\frac{m}{2\hbar (t_2-t_1)}\right)(x_1x_2)^{1/2}\exp\left[\frac{im(x_1^2+x_2^2)}{2\hbar (t_2-t_1)}\right]H_{\gamma}^{(2)}\left(\frac{mx_1x_2}{\hbar (t_2-t_1)}\right)
\end{equation}
where $H_\gamma^{(2)}(z)$ is the Hankel function of the second kind of order
\begin{equation}
\gamma = \sqrt{\frac{1}{4} - \frac{2 m\tilde{\alpha}}{\hbar^2}} 
= ia. 
\end{equation}
which is a dimensionless constant and we have substituted for $\tilde{\alpha}$ from~\eq{potential}. The effective path defined in \eq{pthkern} is given by the integral,
\begin{equation}
X(t) =  \frac{\langle x_2,t_2|\hat{x}(t)|x_1,t_1\rangle}{\langle x_2,t_2|x_1,t_1\rangle} = \frac{1}{K(x_2,t_2|x_1,t_1)}\int_{-\infty}^{\infty} \mathrm{d}x\,x\,K(x_2,t_2|x,t)\,K(x,t|x_1,t_1).
\label{XT}
\end{equation}
Substituting the kernel from \eq{kernel}, we get,
\begin{align}
X(t) = &\lambda\:\exp\left[\frac{- i\pi}{2}(ia+1)\right]\exp\left[\frac{im}{2\hbar}\left(\frac{x_2^2}{t_2 - t}+\frac{x_1^2}{t - t_1} - \frac{(x_1^2 + x_2^2)}{t_2 - t_1}\right)\right]\left[H_{ia}^{(2)} \left(\frac{m x_1 x_2}{\hbar (t_2 - t_1)}\right)\right]^{-1}\nonumber\\
&\int_{-\infty}^{\infty}\,\mathrm{d}x\, x^2 e^{i\lambda x^2} H_{ia}^{(2)}(p x) H_{ia}^{(2)}(q x)
\end{align}
where we have defined,
\begin{equation}
\lambda \equiv \frac{m(t_2 - t_1)}{2\hbar(t_2 - t)(t - t_1)} \hspace{2pt}\hbox{,} \hspace{14pt}
p \equiv \frac{m x_1}{\hbar (t - t_1)}\hspace{7pt}\hbox{and}\hspace{7.5pt} 
q \equiv \frac{m x_2}{\hbar (t_2 -t)}. 
\end{equation}
Note that $\lambda$ has the dimension of inverse length squared while $p$ and $q$ both have dimensions of inverse length. Since the interesting physics takes place when a particle crosses the singularity at the origin, $x=0$, we will take $x_1 = -\epsilon$ at $t_1 = 0$ and $x_2 = \epsilon$ as $t_2\rightarrow\infty$  with limit $\epsilon \rightarrow 0^+$ taken eventually so that the particle has to cross from left to right in the late-time limit. To begin with it is convenient to keep $t_1$ and $t_2$  arbitrary and take the limit at the end of the calculation. Under these conditions, the effective path becomes
\begin{align}
\label{rawX}
X(t) &= \lambda e^{- i \frac{\pi}{2}(i a + 1)} \left[H_{ia}^{(2)}\left(\frac{-m\epsilon^2}{\hbar (t_2 - t_1)}\right)\right]^{-1}\int_{-\infty}^{\infty}\mathrm{d}x\, x^2 e^ {i\lambda x^2}H_{ia}^{(2)}\left( \frac{-m \epsilon x}{\hbar (t - t_1)}\right) H_{ia}^{(2)}\left(\frac{m\epsilon x}{\hbar (t_2 - t)}\right)\nonumber\\
\end{align}
Unfortunately, the integral in the above expression cannot be evaluated exactly in closed from but we can calculate it under the limit $\epsilon\rightarrow0^+ $ as follows. We first express the Hankel functions in the integrand in terms of the Bessel functions which reduces the integral to the form,
\begin{align}
\label{integral}
I = &\int_{-\infty}^{\infty}\mathrm{d}x\, x^2 e^ {i\lambda x^2}H_{ia}^{(2)}\left( \frac{-m\epsilon x}{\hbar (t - t_1)}\right) H_{ia}^{(2)}\left(\frac{m\epsilon x}{\hbar (t_2 - t)}\right)\nonumber\\
= &\left(1-\coth \pi a\right)^2\int_{-\infty}^\infty\mathrm{d}x\;x^2e^{i\lambda x^2}J_{ia}(px)J_{ia}(qx) + \frac{1}{\sinh^2\pi a}\int_{-\infty}^\infty\mathrm{d}x\;x^2e^{i\lambda x^2}J_{-ia}(px)J_{-ia}(qx) \nonumber\\
&+\frac{\left(1-\coth \pi a\right)}{\sinh\pi a}\int_{-\infty}^\infty\mathrm{d}x\;x^2e^{i\lambda x^2}\left(J_{ia}(px)J_{-ia}(qx)+J_{-ia}(px)J_{ia}(qx)\right).
\end{align}
Now we can use the following identity (see~\cite{grad}),
\begin{align}
\int_0^\infty \mathrm{d}x\;x^{\lambda+1}e^{-\alpha x^2}J_\mu(\beta x)J_\nu(\gamma x) &= \frac{\beta^\mu \gamma^\nu\alpha^{-(\mu+\nu+\lambda+2)/2}}{2^{\nu+\mu+1}\Gamma(\nu+1)}\sum_{m=0}^\infty\frac{\Gamma(m+\frac{1}{2}(\nu+\mu+\lambda+2))}{\Gamma(m+\mu+1)\Gamma(m+1)}\left(\frac{-\beta^2}{4\alpha}\right)^m\nonumber\\ 
&\hspace{15.5pt}F(-m,-\mu-m;\nu+1;\frac{\gamma^2}{\beta^2})
\end{align}
and evaluate the integral in the  limit of $\epsilon\to0^+$ (see~\app{appintegralcalc} for details). In the same limit the Hankel function in the denominator can be approximated by: 
\begin{equation}
H_{ia}^{(2)}(z) \approx \frac{i2^{-ia}e^{-\pi a}\Gamma(-ia)z^{ia}}{\pi} + \frac{i 2^{ia}\Gamma(ia)z^{-ia}}{\pi}
\end{equation}
 With these manipulations the effective path can be expressed as,
\begin{equation}
\label{bhX}
X(t) = - i\lambda\,e^{\pi a/2}\,\frac{I}{D}
\end{equation}
where
\begin{align}
I &=\left\{-\frac{e^{\pi a/2}}{2\pi^2} 2^{-ia}\left[\Gamma(-ia)\right]^2\left(\frac{m\epsilon^2}{\hbar(t_2-t_1)}\right)^{ia}\left(1 + e^{-2\pi a}\right)(-i\lambda)^{-3/2}(ia+1/2)\Gamma(ia+1/2)\right.\nonumber
\\&- \left. \frac{e^{-\pi a/2}\,2^{ia}\left[\Gamma(ia)\right]^2}{\pi^2}\left(\frac{m\epsilon^2}{\hbar(t_2-t_1)}\right)^{-ia}\left(1 + e^{-2\pi a}\right)(-i\lambda)^{-3/2}(-ia+1/2)\Gamma(-ia+1/2)\right.\nonumber\\
& \left. -\frac{e^{\pi a}\sqrt{\pi}}{2\pi a\sinh\pi a}(-i\lambda)^{-3/2}\left[e^{\pi a}\left(\frac{t_2-t}{t-t_1}\right)^{ia}+e^{-\pi a}\left(\frac{t_2-t}{t-t_1}\right)^{-ia}\right]\right\}
\end{align}
and
\begin{equation}
D = \left[\frac{i2^{-ia}\Gamma(-ia)}{\pi}\left(\frac{m\epsilon^2}{\hbar (t_2-t_1)}\right)^{ia} + \frac{i2^{ia}\Gamma(ia)}{\pi}\left(\frac{m\epsilon^2}{\hbar (t_2-t_1)}\right)^{-ia} e^{-\pi a}\right].
\end{equation}

Based on our previous analysis of forced harmonic oscillator in~\sec{secho}, we would suspect $|X|^2$ to contain information about the analogue of particle creation in a quantum theory. It is obvious that $|X|^2$ arising from the above  expression will be quite complicated partially due to the fact that it is evaluated for finite time and space interval. To understand the physical significance of this quantity it is again useful to take the limit of $t_2 \rightarrow \infty$ with $t_1=0$ and $\epsilon\rightarrow 0^+$. In this limit, one can ignore transient terms which oscillate rapidly and obtain a simpler expression for $|X|^2$. Somewhat tedious but straight forward algebra (see~\app{appmodxsqrcalc}) yields an interesting final result: We find that $|X|^2$ increases linearly with time allowing us to define a constant, finite, rate which itself takes a very suggestive form as:
\begin{equation}
\label{modxsqrresult}
\frac{\mathrm{d}|X(t)|^2}{\mathrm{d}t} = \left(\frac{4\hbar}{ma}\right)\left(a^2+\frac{1}{4}\right)\left[N +\frac{1}{2}\right]
\end{equation}
where 
\begin{equation}
N = \frac{1}{e^{2\pi a} - 1}
\end{equation}
has the form of a Planckian spectrum of particles. If one thinks of $d|X|^2/dt$ as the rate of production of particles, then it is rather curious that we have a thermal radiation term related to a parameter in the potential, $a$. Obviously, in this particular quantum mechanical example, this result has no physical interpretation but we will next show how this result connects up with radiation from a horizon.
 
\subsection{Quantum mechanics of the scalar field near the horizon}
It turns out that the problem of a scalar field near a black hole spacetime (more generally in any spacetime with a horizon when we consider the Rindler limit of the horizon) can be reduced to that of a quantum mechanical particle in an inverse square potential. In that context, the $d|X|^2/dt$ can be thought of as rate of production of particles by the horizon and the mathematical result obtained above acquires a physical meaning. 

We shall first briefly sketch how the problem of a scalar field near a horizon can be mapped to a quantum mechanical problem of a particle in an inverse-square potential  \cite{TPSreeni}. Consider a scalar field in a 1+1 spacetime with the metric 
\begin{equation}
\label{metric}
ds^2 = B(r)dt^2 - B^{-1}(r)dr^2
\end{equation}
where $B(r)$ has a simple zero  at $r=r_0$ with $B'(r) = dB/dr$ being finite and nonzero at $r_0$. (We will work with (1+1) dimensional system since it captures all the essential physics.) The vanishing of $B(r)$ at point $r=r_0$ indicates the presence of a horizon. Near the horizon, we can expand $B(r)$ as 
\begin{equation}
B(r) = B'(r_0)(r - r_0) + \mathcal{O}[(r - r_0)^2] \approx B'(r_0)(r - r_0).
\end{equation}
Note that in the Schwarzschild case, $B'(r_0) = r_0^{-2}$ with $r_0 = 2M$ as the Schwarzschild radius. The field equation for the scalar field $\Phi(t,r)$,
\begin{equation}
\left( \Box + \frac{m_0^2 c^2}{\hbar^2}\right) \Phi = 0 
\end{equation}
when written for the metric in \eq{metric} becomes
\begin{equation}
c^{-2}B(r)^{-1}\,\partial^2_t\Phi - \partial_r\left(B(r)\partial_r\Phi\right) = - m_0^2c^2\hbar^{-2}\,\Phi. 
\end{equation}
We substitute the following ansatz for $\Phi$ in the above equation,
\begin{equation}
\Phi (r,t) = e^{-i \omega t}\frac{\psi(r)}{\sqrt{B(r)}}
\end{equation} 
and find that $\psi(r)$ satisfies the equation
\begin{equation}
\label{effschr}
- \frac{\hbar^2}{2}\,\frac{d^2\psi(r)}{dr^2} - \frac{\alpha}{(r - r_0)^2}\,\psi(r) = 0    
\end{equation}
where $\alpha = \hbar^2\omega^2/2 c^2[B'(r_0)]^2$ near the horizon (Note that in the near-horizon limit, the term with $m_0$ does not contribute in the leading order). For the Schwarzschild metric, $\alpha = \hbar^2\omega^2r_0^2/2c^2$ hence we see that $\alpha$ has dimensions of $\hbar^2$, as it should.  With $ x = (r - r_0) $, and mass, $m$ put in, this equation is same as the Schr$\ddot{\mathrm{o}}$dinger equation for a particle in an inverse square potential, $-\tilde{\alpha}/x^2$, 
\begin{equation}
- \frac{\hbar^2}{2m}\,\frac{d^2\psi(x)}{dx^2} - \frac{\tilde{\alpha}}{x^2}\,\psi(x) = \mathcal{E}\psi(x)
\end{equation}
where $\tilde{\alpha} =\alpha/m$ and we take the energy eigenvalue $\mathcal{E}\,\rightarrow\,0$ at the end of the calculations.  Thus the problem of scalar field in Schwarzschild background is equivalent to quantum mechanics of a particle in an inverse square potential near the origin. 


\subsection{Horizon thermodynamics}
\label{sechorizon}
With the problem of a scalar field in the Schwarzschild background reduced to an effective quantum mechanical problem  in an inverse square potential, we can identify the parameters of potential in the two situations.
\begin{equation}
V(x) = \frac{\hbar^2}{2m}\left(a^2 + \frac{1}{4}\right)\frac{1}{x^2} = -\frac{\hbar^2\omega^2r_0^2}{2m c^2}\frac{1}{x^2} 
\end{equation}
which gives for $a$,
\begin{equation}
a = \left(\frac{\omega^2r_0^2}{c^2}-\frac{1}{4}\right)^{1/2}\approx \frac{\omega r_0}{c}.
\end{equation}
in the high-frequency limit. In this case,
substituting for $a$ in our expression for $d|X|^2/dt$ obtained earlier gives,
\begin{equation}
\frac{\mathrm{d}|X(t)|^2}{\mathrm{d}t} = \frac{8GM}{mc^3}\left[\hbar\omega\left(N +\frac{1}{2}\right)\right]
\end{equation}
where
\begin{equation}
N = \frac{1}{e^{2\pi a} - 1} = \frac{1}{e^{\hbar\omega/k_BT}-1}
\end{equation}
with the temperature being given by
\begin{equation}
T = \frac{\hbar c^3}{4\pi GMk_B}=\frac{1}{4\pi M}
\end{equation}
where the second result is valid in natural units. (The time asymmetry in our boundary conditions, $t_1=0, t_2\to\infty$ makes it meaningful to treat $d|X|^2/dt$ as a rate.)  
The expression $N$, of course, represents a Planckian spectrum of particles at temperature $T$ and the part $N+(1/2)$ correctly describes the Planckian energy density of a cavity at temperature $T$ along with the zero point contribution. The temperature $T=1/4\pi M$ however is twice the usual temperature associated with black holes. This feature is well-known in the tunneling derivation of black hole temperature and has been extensively discussed in the literature \cite{factortwo}. While the topic is still somewhat controversial, the origin of this extra factor is attributed to using singular coordinates at horizon \cite{tunnelingreview}. Since we have started with Schwarzschild coordinates (which \textit{are} ill-defined at the horizon) it is probably natural that we get this result.

Thus the effective path approach seems to capture the Planck spectrum (with the temperature off by factor 2 which occurs in some other tunneling computations as well) along with zero-point energy. So
the squared modulus of $X(t)$ does contain information related to the production of particles, this time in a fairly non-trivial setting. 


\section{Conclusions}\label{conclu}

The concept of effective action is a well-known technique that is used in the literature 
to study  various aspects of quantum field theories in classical backgrounds. The effective action is in general complex and its real and imaginary parts contain information about the vacuum polarization and particle production. In using the effective action to describe the back reaction effects, one usually uses the real part of the effective action and discards the imaginary part in order to obtain real equations of motion. 

On the other hand if we retain both real and imaginary parts of the effective action and obtain the equations of motion, then the solutions will be --- in general --- complex. Because the imaginary part of the effective action contains information about the particle production, it seems likely that the solutions to the complex effective action will give us a handle to explore particle production. This motivates us to study a quantum effective path $X(t)$ (in the context of QM) which is a solution to $\delta \Gamma[X]=0$. Fortunately, this $X(t)$ can be expressed as an integral over the path integral kernel and hence can be evaluated, in principle, if the kernel is known.
 
In practice, the calculation turns out to be quite complicated. To gather a preliminary insight we studied two important examples in this paper. First one is the case of a forced harmonic oscillator in which we could directly link the complex effective path to particle production in the asymptotic limit. The imaginary part of the effective path is generated 
 solely by the non-zero Fourier mode of the external source at the natural frequency of the oscillator. The modulus square of the complex effective path gave the particle production rate in the system. (We also found that the complex effective path obtained in this case also provided a nice interpretation to a quantity which was purely a mathematical construction by Landau used in the case of forced harmonic oscillator.)

The second case we studied was that of an attractive, inverse-square potential. It was known from previous work that the problem of a scalar field in a spacetime with a horizon (in which the near-horizon geometry can be approximated as Rindler geometry) can be mapped to the Schr$\mathrm{\ddot{o}}$dinger problem in an inverse square potential. We expect the emission of particles by the black hole to get mapped to propagation of particle through the singularity at the origin in the equivalent Schr$\mathrm{\ddot{o}}$dinger problem, even though there are no time-dependent sources. In this case the modulus square of the effective path can be interpreted as a rate of emission of particles. This expression correctly gives  the Planckian distribution along with the zero point contribution  for the Hawking radiation. The temperature of the Planckian distribution turns out to be  $T = \hbar/4\pi M$ which is twice the standard value for Hawking temperature. 
This factor of two discrepency has been noticed in the literature previously and arises when one uses coordinates which are singular at the horizon and hence is probably understandable.

Finally we would like to make some comments regarding the nature of the
potential ($V(x)\propto -x^{-2}$) considered in the last section and the existence of imaginary
part in the trajectory. The first example we studied
in section 3 has an explicitly time-dependant external source and hence it
is not surprising that we encounter particle production and an imaginary part to
the classical trajectory. In the case of an inverse square potential
there is no explicit time-dependance but we still obtain an imaginary part to the trajectory.
In fact, the wave equation for a scalar field in the black hole spacetime --- which contains the physics of black hole evaporation --- does get mapped to such a static
potential. It is, however, known from previous work that particle creation
can occur even in the absence of explicit time-dependance. A 
well studied example of this is the Schwinger effect in which one obtains
a steady particle production in the presence of static electric field. In
this particular case the wave equation can be mapped to an inverted harmonic
oscillator \cite{TPSreeni} for which the Hamiltonian is unbounded. In a way, it is this
singular behavior of the Hamiltonian which leads to the particle
production. (In contrast, the wave equation in the presence of a constant magnetic field
gets mapped to a normal harmonic oscillator with a bounded Hamiltonian and
--- as expected --- one does not have any particle creation in a constant magnetic
field.) The current situation is very similar: The wave equation in the black hole spacetime gets mapped to an inverse square potential and it is well-known that this
potential leads to a Hamiltonian which is not hermitian. In the previous
work \cite{TPSreeni} which connects up  particle production in blackhole spacetime 
 with the inverse square potential one crucially used the singular
structure of the potential (and an integration around a singularity in
complex plane) to obtain the result. This work has established the essential
connection between the singular nature of the Hamiltonian in these potentials (both in the
context of black hole spacetimes as well as in the context of constant electric field) and the
production of particles. We,  therefore, believe that path integral
formalism studied in this paper leads to complex trajectories for
essentially the same reason viz. that the Hamiltonian is non-Hermitian. It would be interesting to investigate this question
further and see whether one can provide a direct and rigorous proof for the the existence of
complex paths for certain class of  Hamiltonians which are unbounded or non-Hermitian.


\section*{Acknowledgements}

SS is supported by a fellowship from the Council of Scientific and Industrial Research (CSIR), India. TP's research is partially supported by the J.C.Bose Research Grant of DST, India. We thank the referee for useful comments.


\appendix{}

\section{Detailed calculation of some results}

\subsection{Path integral kernel for an inverse square potential}
\label{appinvsqrkernel}
The path integral kernel is defined by,
\begin{equation}
K(x_2,T|x_1,0) = \int\mathcal{D}x(t)\,e^{\frac{i}{\hbar}[\int_0^t\mathrm{d}t\,(\frac{1}{2}m\dot{x}^2 - \alpha x^{-2})]} 
\end{equation}
To evaluate the kernel, we use the perturbative series expansion, which gives
\begin{eqnarray}
K(x_2,T|x_1,0) &=& K_0(x_2,t_2|x_1,t_1) + \sum_{n\,=\,1}^\infty\left(\frac{-i\alpha}{\hbar}\right)^n\int_0^T\mathrm{d}t_n\int_0^{t_n}\mathrm{d}t_{n-1}\cdot\nonumber\\
               & & \cdot\cdot\int_0^{t_2}\mathrm{d}t_1 \int\prod_{j=1}^{n}\frac{\mathrm{d}x_j}{x_j^2}\left(\prod_{j=0}^nK_0(x_{j + 1},t_{j + 1}|x_j,t_j)\right) 
\end{eqnarray}
where $K_0(x_2,t_2|x_1,t_1)$ is the free particle kernel. Introducing
\begin{equation}
G(x_2,x_1;E) \equiv \int_0^\infty \mathrm{d}T e^{-\frac{i}{\hbar}ET}K(x_2,T|x_1,0)
\end{equation}
we have,
\begin{eqnarray}
G(x_2,x_1;E) &=& G_0(x_2,x_1;E) + \sum_{n\,=\,1}^\infty\left(\frac{-i\alpha}{\hbar}\right)^n \int\prod_{j=1}^{n}\frac{\mathrm{d}x_j}{x_j^2}\left(\prod_{j=0}^nG_0(x_{j + 1},x_j;E)\right)\nonumber\\
             &=& G_0(x_2,x_1;E) + \sum_{n\,=\,1}^\infty \left(\frac{-i\alpha}{\hbar}\right)^n G_n  
\end{eqnarray}
We need to sum up the above series so as to get the closed form for the Kernel. To do this we employ a trick \cite{trick} in which we first express the free particle propagator, $G_0(x_2,x_1;E)$ in terms of the Hankel functions and then use their orthogonality relation to evaluate the $n^{th}$ order product, $G_n$.
\begin{align}
\label{freegreenfunc}
G_0(x_2,x_1;E) &\equiv \frac{1}{i}\left(\frac{m}{2E}\right)^{1/2}e^{ik|x_2 - x_1|}\nonumber\\
&=\left(\frac{m\pi}{i2\hbar}\right)(x_1x_2)^{1/2}\,H_{1/2}^{(1)}(kx_>)H_{1/2}^{(2)}(kx_<)\nonumber\\
&= \left(\frac{m}{i\hbar}\right)(x_1x_2)^{1/2}\int_0^\infty\mathrm{d}\nu\,\frac{\nu\sinh(\nu\pi)}{\nu^2 + 1/4}H_{i\nu}^{(1)}(kx_2)H_{i\nu}^{(1)*}(kx_1)            
\end{align}
where $k = \sqrt{\frac{2 m E}{\hbar^2}}$. Upon inserting the expression for $G_0$, the n-fold integrations can be performed using the orthogonality relation
\begin{equation}
\int_0^\infty \frac{\mathrm{d}x}{x}\,H_{i\nu}^{(1)*}(kx)H_{i\nu^\prime}^{(1)}(kx) = \frac{2\delta(\nu-\nu^\prime)}{\nu\sinh\,\nu\pi}\nonumber
\end{equation}
and we obtain
\begin{equation}
G_n(x_2,x_1;E) = \left(\frac{2m}{i\hbar}\right)^{n + 1}\frac{(x_1x_2)^{1/2}}{2} \int_0^\infty\mathrm{d}\nu\,\frac{\nu\sinh(\nu\pi)}{\nu^2 + 1/4}H_{i\nu}^{(1)}(kx_2)H_{i\nu}^{(1)*}(kx_1)
\end{equation}
We substitute this expression for $G_n$ and sum the resulting geometric series to get
\begin{equation}
G(x_2,x_1;E) = \left(\frac{m}{i\hbar}\right)(x_1x_2)^{1/2}\int_0^\infty\mathrm{d}\nu\,\frac{\nu\sinh(\nu\pi)}{\nu^2 + 1/4 + \frac{2m\alpha}{\hbar^2} }H_{i\nu}^{(1)}(kx_2)H_{i\nu}^{(1)*}(kx_1)
\end{equation}
which is the exact expression and is similar to the free Green function in~\eq{freegreenfunc} with an addition to the denominator of the integrand. Noting that the the free particle Kernel can be written as,
\begin{eqnarray}
K_0 &=& \left(\frac{m}{2\pi i\hbar T}\right)^{1/2}\exp\left[\frac{im}{2\hbar T}(x_2 - x_1)^2\right]\nonumber\\
    &=& e^{-\frac{i3\pi}{4}}\left(\frac{m}{2\hbar T}\right)(x_1x_2)^{1/2}\exp\left[\frac{im(x_1^2+x_2^2)}{2\hbar T}\right]H_{1/2}^{(2)}\left(\frac{mx_1x_2}{\hbar T}\right) 
\end{eqnarray}
We can obtain the Kernel for the problem by suitable replacements in the free particle Kernel due to modified denominator in (59) from (57) as
\begin{equation}
K(t_2,x_2|t_1,x_1) = e^{-\frac{1}{2}i\pi(\gamma + 1)}\left(\frac{m}{2\hbar T}\right)(x_1x_2)^{1/2}\exp\left[\frac{im(x_1^2+x_2^2)}{2\hbar T}\right]H_{\gamma}^{(2)}\left(\frac{mx_1x_2}{\hbar T}\right)
\end{equation}
where 
\begin{equation}
\gamma = \sqrt{\frac{1}{4} + \frac{2 m\alpha}{\hbar^2}} 
\end{equation}
When $\alpha = 0$, we have $\gamma=1/2$ and the expression reduces to the free-particle Kernel.

\subsection{Evaluation of the integral in~\eq{integral}}
\label{appintegralcalc}
We have,
\begin{align}
I = &\int_{-\infty}^{\infty}\mathrm{d}x\, x^2 e^ {i\lambda x^2}H_{ia}^{(2)}\left( \frac{-m\epsilon x}{\hbar (t - t_1)}\right) H_{ia}^{(2)}\left(\frac{m\epsilon x}{\hbar (t_2 - t)}\right)\nonumber\\
= &\left(1-\coth \pi a\right)^2\int_{-\infty}^\infty\mathrm{d}x\;x^2e^{i\lambda x^2}J_{ia}(px)J_{ia}(qx) + \frac{1}{\sinh^2\pi a}\int_{-\infty}^\infty\mathrm{d}x\;x^2e^{i\lambda x^2}J_{-ia}(px)J_{-ia}(qx) \nonumber\\
&+\frac{\left(1-\coth \pi a\right)}{\sinh\pi a}\int_{-\infty}^\infty\mathrm{d}x\;x^2e^{i\lambda x^2}\left(J_{ia}(px)J_{-ia}(qx)+J_{-ia}(px)J_{ia}(qx)\right).
\end{align}
Using the following identity(see~\cite{grad}),
\begin{align}
\int_0^\infty \mathrm{d}x\;x^{\lambda+1}e^{-\alpha x^2}J_\mu(\beta x)J_\nu(\gamma x) &= \frac{\beta^\mu \gamma^\nu\alpha^{-(\mu+\nu+\lambda+2)/2}}{2^{\nu+\mu+1}\Gamma(\nu+1)}\sum_{m=0}^\infty\frac{\Gamma(m+\frac{1}{2}(\nu+\mu+\lambda+2))}{\Gamma(m+\mu+1)\Gamma(m+1)}\left(\frac{-\beta^2}{4\alpha}\right)^m\nonumber\\ 
&\hspace{15.5pt}F(-m,-\mu-m;\nu+1;\frac{\gamma^2}{\beta^2})
\end{align}
the three integrals in $I$ can be evaluated in the following manner. 
\begin{align}
I_1 &= \left(1-\coth \pi a\right)^2 \int_{-\infty}^\infty \mathrm{d}x\;x^2e^{i\lambda x^2}J_{ia}(px)J_{ia}(qx) \nonumber\\
&= \frac{e^{-2\pi a}}{(\sinh \pi a)^2}\left(1+e^{2\pi a}\right) \int_0^\infty \mathrm{d}x\;x^2e^{i\lambda x^2}J_{ia}(px)J_{ia}(qx)\nonumber\\
&= \frac{\left(1+e^{-2\pi a}\right)}{(\sinh \pi a)^2}\frac{(pq)^{ia}(-i\lambda)^{-ia-3/2}}{2^{2ia+1}\Gamma(ia +1)}\sum_{n=0}^{\infty}\frac{\Gamma(n+ia+3/2)}{n!\Gamma(n+ia+1)}\left(\frac{p^2}{4i\lambda}\right)^n F\left(-n,-ia-n;ia+1;\frac{q^2}{p^2}\right)\nonumber\\
&= \frac{e^{\pi a/2 }\left(1+e^{-2\pi a}\right)}{2(\sinh \pi a)^2}\left(\frac{m\epsilon^2}{\hbar (t_2-t_1)}\right)^{ia}\frac{(-i\lambda)^{-3/2}2^{-ia}}{\Gamma(ia +1)}\sum_{n=0}^{\infty}\frac{\Gamma(n+ia+3/2)}{n!\Gamma(n+ia+1)}\left(\frac{(t_2-t_1)m\epsilon^2}{2\hbar i (t-t_1)(t_2-t)}\right)^n\nonumber\\
&\hspace{25pt}F\left(-n,-ia-n;ia+1;\left(\frac{t-t_1}{t_2-t}\right)^2\right).\nonumber
\end{align}
This expression cannot be simplified further in the general case. However, we are interested in the $\epsilon\rightarrow 0$ limit when only the $n=0$ term contributes and the expression reduces to:
\begin{align}
I_1&= \frac{e^{\pi a/2 }\left(1+e^{-2\pi a}\right)}{2(\sinh \pi a)^2}\frac{(-i\lambda)^{-3/2}2^{-ia}}{[\Gamma(ia +1)]^2}\Gamma(ia+3/2)\left(\frac{m\epsilon^2}{\hbar (t_2-t_1)}\right)^{ia}\nonumber\\
&= -\frac{ e^{\pi a/2 }\left(1+e^{-2\pi a}\right)}{2\pi^2}(-i\lambda)^{-3/2}2^{-ia}[\Gamma(-ia)]^2\Gamma(ia+3/2)\left(\frac{m\epsilon^2}{\hbar (t_2-t_1)}\right)^{ia}
\end{align}
The integral $I_2$ is same as $I_1$ with $a\rightarrow -a$. Therefore,
\begin{align}
I_2 &= \frac{1}{(\sinh\pi a)^2} \int_{-\infty}^\infty \mathrm{d}x\;x^2e^{i\lambda x^2}J_{-ia}(px)J_{-ia}(qx)\nonumber\\ 
      &= e^{-2\pi a}I_1(a\rightarrow -a)\nonumber\\
&= -e^{-2\pi a}\frac{e^{-\pi a/2}\left(1+e^{2\pi a}\right)}{2\pi^2}(-i\lambda)^{-3/2}2^{ia}[\Gamma(ia)]^2\Gamma(-ia+3/2)\left(\frac{m\epsilon^2}{\hbar (t_2-t_1)}\right)^{-ia}\nonumber\\
&= -\frac{ e^{-\pi a/2 }\left(1+e^{-2\pi a}\right)}{2\pi^2}(-i\lambda)^{-3/2}2^{ia}[\Gamma(ia)]^2\Gamma(-ia+3/2)\left(\frac{m\epsilon^2}{\hbar (t_2-t_1)}\right)^{-ia}.
\end{align}
Similarly, we can evaluate the third integral as well,
 \begin{align}
I_3 &= \frac{(1-\coth\pi a )}{\sinh\pi a} \int_{-\infty}^\infty \mathrm{d}x\;x^2e^{i\lambda x^2}[J_{ia}(px)J_{-ia}(qx) + J_{ia}(qx)J_{-ia}(px)] \nonumber\\
&= -\frac{2e^{-\pi a}}{(\sinh \pi a)^2}\int_0^\infty \mathrm{d}x\;x^2e^{i\lambda x^2}[J_{ia}(px)J_{-ia}(qx) + J_{ia}(qx)J_{-ia}(px)] \nonumber\\
&= \frac{-2e^{-\pi a}}{(\sinh \pi a)^2}\left[\frac{p^{ia}q^{-ia}(-i\lambda)^{-\frac{3}{2}}}{2\Gamma(1-ia)}\sum_{n=0}^{\infty}\frac{\Gamma(n+\frac{3}{2})}{n!\Gamma(n+ia+1)}\left(\frac{p^2}{4i\lambda}\right)^n F\left(-n,-ia-n;-ia+1;\frac{q^2}{p^2}\right) + (a\rightarrow -a)\right] \nonumber\\
&= -\frac{e^{-\pi a}\sqrt{\pi}}{2\pi a\sinh\pi a}(-i\lambda)^{-3/2}\left\{e^{\pi a}\left(\frac{t_2-t}{t-t_1}\right)^{ia}+e^{-\pi a}\left(\frac{t_2-t}{t-t_1}\right)^{-ia}\right\}.
\end{align}
 Combining the results,
\begin{align}
\label{N}
I &= I_1 + I_2 +I_3\nonumber\\
 &=\left\{-\frac{e^{\pi a/2}}{2\pi^2} 2^{-ia}\left[\Gamma(-ia)\right]^2\left(\frac{m\epsilon^2}{\hbar(t_2-t_1)}\right)^{ia}\left(1 + e^{-2\pi a}\right)(-i\lambda)^{-3/2}(ia+1/2)\Gamma(ia+1/2)\right.\nonumber
\\&- \left. \frac{e^{-\pi a/2}\,2^{ia}\left[\Gamma(ia)\right]^2}{\pi^2}\left(\frac{m\epsilon^2}{\hbar(t_2-t_1)}\right)^{-ia}\left(1 + e^{-2\pi a}\right)(-i\lambda)^{-3/2}(-ia+1/2)\Gamma(-ia+1/2)\right.\nonumber\\
& \left. -\frac{e^{-\pi a}\sqrt{\pi}}{2\pi a\sinh\pi a}(-i\lambda)^{-3/2}\left[e^{\pi a}\left(\frac{t_2-t}{t-t_1}\right)^{ia}+e^{-\pi a}\left(\frac{t_2-t}{t-t_1}\right)^{-ia}\right]\right\}.
\end{align}

\subsection{Evaluation of $|X|^2$ in the \eq{modxsqrresult}}
\label{appmodxsqrcalc}
We have the effective path,
\begin{equation}
X(t) = - i\lambda\,e^{\pi a/2}\frac{I}{D}
\end{equation}
where $I$ is given by \eq{N} above and
\begin{equation}
D = \left[\frac{i2^{-ia}\Gamma(-ia)}{\pi}\left(\frac{m\epsilon^2}{\hbar (t_2-t_1)}\right)^{ia} + \frac{i2^{ia}\Gamma(ia)}{\pi}\left(\frac{m\epsilon^2}{\hbar (t_2-t_1)}\right)^{-ia} e^{-\pi a}\right].
\end{equation}

In general, $|X|^2$ arising from the above  expression will be quite complicated. But, working in the limit of $t_2 \rightarrow \infty$ with $t_1=0$ and $\epsilon\rightarrow 0^+$, we will be able to extract a meaningful result. To see this, first note that  $I_1$, $I_2$ and $D$ can be written as 
\begin{align}
I_1 &=  -  \frac{e^{\pi a/2}(1+e^{-2\pi a})}{2\pi a(-i\lambda)^{3/2}\sinh\pi a}\sqrt{\frac{\pi}{\cosh \pi a}}\sqrt{a^2+\frac{1}{4}}\,\exp\left[i(2\theta+\psi+\phi)+ia\ln\frac{m\epsilon^2}{2\hbar (t_2-t_1)}\right]\nonumber\\
I_2 &=  -  \frac{e^{-\pi a/2}(1+e^{-2\pi a})}{2\pi a(-i\lambda)^{3/2}\sinh\pi a}\sqrt{\frac{\pi}{\cosh \pi a}}\sqrt{a^2+\frac{1}{4}}\,\exp\left[-i(2\theta+\psi+\phi)-ia\ln\frac{m\epsilon^2}{2\hbar (t_2-t_1)}\right]\nonumber\\
D &= \frac{i}{\pi}\sqrt{\frac{\pi}{a\sinh\pi a}}\left\{\exp\left[i\theta+ia\ln\frac{m\epsilon^2}{2\hbar (t_2-t_1)}\right]+e^{-\pi a}\exp\left[-i\theta-ia\ln\frac{m\epsilon^2}{2\hbar (t_2-t_1)}\right]\right\}
\end{align}
where,
\begin{align}
\theta = \arg [\Gamma(-ia)], \hspace{5pt} \phi = \arg[\Gamma(ia+1/2)]\hspace{5pt}\mathrm{and}\hspace{2pt}\psi = \arg[ia+1/2] 
\end{align}
Then, one sees immediately, that
\begin{align}
\frac{I_1+I_2}{D} =& \frac{ie^{\pi a/2}(1+e^{-2\pi a})}{2a(-i\lambda)^{3/2}\sinh\pi a}\sqrt{\frac{a\sinh\pi a}{\cosh\pi a}}\sqrt{a^2+\frac{1}{4}}\hspace{3pt}e^{i(\theta+\phi+\psi)}\nonumber\\
&\hspace{5pt}\frac{1+e^{-\pi a}\exp\left[-i(4\theta+2\phi+2\psi) - 2ia\ln\left(m\epsilon^2/2\hbar (t_2-t_1)\right)\right]}{1+e^{-\pi a}\exp\left[-i2\theta - 2ia\ln\left(m\epsilon^2/2\hbar (t_2-t_1)\right)\right]}
\end{align}
Similarly,
\begin{equation}
\frac{I_3}{D} = \frac{ie^{-\pi a}}{2(-i\lambda)^{-3/2}\sqrt{a\sinh\pi a}}\frac{e^{\pi a}\exp\left[ia\ln \frac{(t_2-t)}{(t-t_1)}\right]+e^{-\pi a}\exp\left[-ia\ln \frac{(t_2-t)}{(t-t_1)}\right]}{\exp\left[i\theta+ia\ln\frac{m\epsilon^2}{2\hbar (t_2-t_1)}\right]+e^{-\pi a}\exp\left[-i\theta-ia\ln\frac{m\epsilon^2}{2\hbar (t_2-t_1)}\right]}
\end{equation}
Imposing the late time condition $t_2 \rightarrow \infty$ with $t_1=0$ and $\epsilon\rightarrow 0$, and ignoring the  oscillatory terms which do not contribute on the average we can simplify this expression. In this limit,  we can neglect the contribution from $I_3/D$ term altogether while pre-factor in the $(I_1+I_2)/D$ term gives
\begin{align}
|X(t)|^2 &= \frac{\lambda^2 e^{2\pi a}(1+e^{-2\pi a})^2}{4a^2\lambda^3\sinh^2\pi a}\frac{a\sinh\pi a}{\cosh\pi a}\left(a^2+\frac{1}{4}\right)\nonumber\\
&= \left(\frac{4\hbar t}{m a}\right)\left[N +\frac{1}{2}\right]\left(a^2+1/4\right)
\end{align}
or 
\begin{equation}
\frac{\mathrm{d}|X(t)|^2}{\mathrm{d}t} = \left(\frac{4\hbar}{m a}\right)\left[N +\frac{1}{2}\right](a^2+1/4)
\end{equation}
where
\begin{equation}
N =  \frac{1}{e^{2\pi a}-1}
\end{equation} 

It is worth mentioning here that if we include the leading transient terms, then the  above expression gets modified by an extra term:
\begin{equation}
\frac{\mathrm{d}|X(t)|^2}{\mathrm{d}t} = \left(\frac{4\hbar}{m a}\right)\left[N +\frac{1}{2}+\sqrt{N(N+1)}\cos \xi\right](a^2+1/4)
\end{equation}
where
\begin{equation}
\xi =  2a\left(\frac{m\epsilon^2}{2\hbar (t_2-t_1)}\right)+ 4\theta+2\phi+2\psi
\label{defxi}
\end{equation} 
The factor $\sqrt{N(N+1)}$ has the physical meaning of the root-mean-square fluctuation of the photons in Planck spectrum (see e.g. \cite{rmsphoton}). Given the large phase in the cosine term (when $\xi\gg 1$), one may say that the relevant term varies rapidly between $-\sqrt{N(N+1)}$ and $\sqrt{N(N+1)}$, matching the magnitude of thermal fluctuations of photons in a bath.
What is probably remarkable is that a similar result was obtained years back \cite{tpplanewavesacc} in a completely different context. In \cite{tpplanewavesacc}, the authors showed that the Fourier transform of a classical plane wave with respect to the Rindler time coordinates leads to a very similar expression with exactly the three terms. It is not obvious why the effective path method should lead to such a result and this similarity is worth investigating. We hope to do this in a future publication. 
  

\section{Effective path for a class of inverse square potentials}  
\label{appeffpathclassofinvsqr}
In general, evaluation of the effective path requires the knowledge of the path integral kernel and tractability of the integral which appears in \eq{pthkern}.  In many cases of interest, algebraic difficulties prevent the analysis of the effective path in an explicit form. Given the fact that it could be a useful tool in probing particle production, we present in this appendix some specific cases in which such a calculation can be performed. We also provide the calculational details for $X(t)$ including the case considered in \cite{brown} since we could not find these details in the literature.\\
\\
The simplest context in which the relevant equations are tractable occurs for a special class of inverse square potentials having the form $V(x) = l(l + 1)\hbar^2(2m)^{-1}x^{-2}$ where $l$ is an integer. Note that this potential has $\alpha > 0$ unlike the case in the previous section and we would like to probe the nature of effective path across the singularity in order to display the tunneling feature via a complex path. For such a case, $\gamma = (l + 1/2)$, and we can using the property of Hankel functions of half-integral orders \cite{grad}, 
\begin{equation}
H_{n+\frac{1}{2}}^{(2)} (z) = i^{n}H_{\frac{1}{2}}^{(2)}(z) \sum_{k = 0}^{n}\frac{(n+k)!}{k!(n-k)!}\frac{1}{(2 i z)^k}, 
\end{equation}
write the generic kernel as
\begin{equation}
K_{\gamma}(x_2,t_2|x_1,t_1) = K_{\frac{1}{2}}(x_2,t_2|x_1,t_1) \sum_{k = 0}^{\gamma-1/2} \frac{(\gamma + k - 1/2)!}{(\gamma - k - 1/2)! k!}\left( \frac{\hbar(t_2-t_1)}{2 i mx_1x_2}\right)^k
\end{equation}
where $K_{1/2}$ is the free particle kernel. The result is a finite series for any particular choice (half-integral) of $\gamma$ and can, in principle, be used to evaluate the effective path for any given value of $\gamma$. 

As an example of the use of this result we will consider the nature of the effective path near the origin along the lines studied in \cite{brown} for a more general case. For this purpose,
 we will see that it suffices to look at two starting simple non-zero values, $l = 1$ and $2$. For the first case, $\alpha = \hbar^2m^{-1}$ and $\gamma = 3/2$, so that we have result which is obtained earlier in \cite{brown}, viz.
\begin{equation}
K_{3/2}(x_2,t_2|x_1,t_1) = \left\{1 - \frac{i\hbar(t_2 -t_1)}{mx_1x_2}\right\}K_{1/2}(x_2,t_2|x_1,t_1).
\end{equation}
The effective trajectory for this case is 
\begin{align}
X_{3/2}(t) &= \frac{1}{K_{3/2}(2|1)}\int\mathrm{d}x K_{3/2}(2|x,t)\,x\,K_{3/2}(x,t|1)\nonumber\\
      &= \frac{1}{K_{3/2}(2|1)}\int\mathrm{d}x K_{1/2}(2|x)\,x\,K_{1/2}(x|1)\left\{1 - \frac{i\hbar (t_2 - t_1)}{m x_1 x_2 x}\bar{x} - \frac{\hbar^2}{m^2}\frac{(t_2 - t)(t - t_1)}{x_1 x_2 x^2} \right\}\nonumber\\ 
      &= \bar{x} + \frac{i\pi\hbar^2\sqrt{m(t_2-t)(t-t_1)(t_2-t_1)}}{(2\pi i\hbar)^{1/2}m(mx_1x_2 - i\hbar(t_2-t_1))}\exp(i\lambda\bar{x}^2)\Phi(\bar{x}(i\lambda)^{1/2})
\end{align}
where 
\begin{equation}
\bar{x} = \frac{x_2(t - t_1) + x_1(t_2 -t)}{(t_2 - t_1)}\nonumber\\
\end{equation}
and $\Phi(x)$ is the probability integral. To study the small $\hbar$ behavior, that is, $\epsilon = \hbar(t_2 - t_1)/(mx_1x_2)\ll1$, we use the properties of $\Phi(x)$ \cite{grad}, and get
\begin{equation}
X_{3/2}(t) = \bar{x}(t) -\hbar^2\frac{(t_2 - t)(t - t_1)}{m^2x_1x_2}[\bar{x}^{-1} - \sqrt{(1/2)i\pi\lambda}\exp(i\lambda\bar{x}^2)] + \cdot\cdot\cdot . 
\end{equation}
To the same order in $\epsilon$ the classical trajectory is given by 
\begin{equation}
\label{xcl}
x_{cl}(t) = \bar{x} - \frac{\hbar^2(t_2 -t)(t-t_1)}{m^2x_1x_2\bar{x}} + \mathrm{O}(\epsilon^3).
\end{equation}
In the limit $\hbar\,\rightarrow\,0$, effective trajectory becomes,
\begin{equation}
X_{3/2}(t) = \bar{x} - \frac{\hbar^2(t_2 -t)(t-t_1)}{m^2x_1x_2}[\bar{x}^{-1} -i\pi\delta(\bar{x})].
\end{equation}
Using
\begin{equation}
\lim_{\eta\,\rightarrow\,0}\,(\bar{x} + i\eta)^{-1} = \bar{x}^{-1} -i\pi\delta(\bar{x})
\end{equation}
we can rewrite this as
\begin{equation}
\label{X3by2}
X_{3/2}(t) = \bar{x} - \frac{\hbar^2(t_2 -t)(t-t_1)}{m^2x_1x_2(\bar{x}+i\eta)}
\end{equation}
We shall now evaluate the effective path for $l=2$ (for which $\gamma = 5/2$) to the same order. In this case the kernel is a series with three terms,
\begin{equation}
K_{5/2}(x_2,t_2|x_1,t_1) = \left\{1 - \frac{i\hbar(t_2 -t_1)}{mx_1x_2} - \frac{3\hbar^2(t_2 -t_1)^2}{m^2x_1^2x_2^2}\right\}K_{1/2}(x_2,t_2|x_1,t_1).
\end{equation}
The calculation for effective path proceeds in the same way although the algebra becomes tedious. Working out the effective path to the same order in $\epsilon$ again shows similar pattern as \eq{X3by2}: 
\begin{equation}
\label{X5by2}
X_{5/2}(t) = \bar{x} + \frac{9\hbar^2(t_2 - t_1)^2}{m^2 x_1^2 x_2^2} \bar{x} - \frac{5\hbar^2}{m^2}\frac{(t_2 -t)(t - t_1)}{x_1x_2(\bar{x}+i\eta)} 
\end{equation}
Note that classically the particle cannot cross the origin and in fact the classical trajectory in \eq{xcl} has a singularity at $\bar x =0$. However, the complex effective trajectories in Eqs.~(\ref{X3by2},\,\ref{X5by2}) are non-singular at $\bar x =0$ since it can move over to the imaginary axis. The trend persists for higher values of $\gamma$ as well, displaying the excursion into the complex plane near the origin. This can be shown quickly in symbolic terms. For any $\gamma > 5/2$, we will have,
\begin{equation}
K_\gamma  (2|1) = K_{1/2}(2|1) \left\{1 + \frac{C_1}{x_1x_2} + \frac{C_2}{(x_1x_2)^2} + \cdot\cdot\cdot + \frac{C_{\gamma - 1/2}}{(x_1x_2)^{\gamma - 1/2}}\right\}
\end{equation} 
Now, the effective path is 
\begin{align}
X_\gamma (t) = \frac{1}{K_\gamma (2|1)} \left[ \int \mathrm{d}x\,x K_{1/2}(2|x,t) K_{1/2}(x,t|1) \left(1 + \frac{A_1}{x x_2} + \frac{A_2}{(x x_2)^2} + \cdot\cdot\cdot + \frac{A_{\gamma - 1/2}}{(x x_2)^{\gamma - 1/2}}\right) \right.  \nonumber\\
\left. \left(1 + \frac{B_1}{x_1x} + \frac{B_2}{(x_1x)^2} + \cdot\cdot\cdot + \frac{B_{\gamma - 1/2}}{(x_1x)^{\gamma - 1/2}}  \right)  \right]\nonumber\\ 
\end{align}
For the first few terms we have, 
\begin{align}
\label{Xgamma}
X_\gamma (t) =  \frac{1}{K_{1/2}(2|1)} \left\{1 - \frac{C_1}{x_1x_2} - \frac{C_2}{(x_1x_2)^2} - \cdot\cdot\cdot \right\} \left[\bar{x} K_{1/2}(2|1) + (A_1/x_2 + B_1/x_1)K_{1/2} (2|1) + \right. \nonumber\\
\left.f(A_1,A_2,B_1,B_2,x_1,x_2)\int\frac{\mathrm{d}x}{x}\, K_{1/2}(2|x,t)K_{1/2}(x,t|1) + \cdot\cdot\cdot \right]
\end{align}
Then, we can easily see that in our limit of $\epsilon\ll 1$,
\begin{equation}
X_\gamma (t) = \mathrm{Re}X_\gamma + i \,\mathrm{Im}X_\gamma  
\end{equation}
where the imaginary part essentially comes from the integral in the \eq{Xgamma} which is the probability integral. Thus the result obtained for $\gamma=3/2$ in \cite{brown} turns out to be true for a much wider class of potentials.



\begin{thebibliography}{25}

\bibitem{schwinger}
J.~Schwinger, Phys.\ Rev.,~{\bf 82}, 664 (1951); 
A.~I.~Nikishov, Zh.\ Eksp.\ Teor.\ Fiz.,~{\bf 57}, 1210 (1969); for a text book discussion see
C.~Itzykson and J.~B.~Zuber, {\sl Quantum Field Theory} (McGraw-Hill, New York, 1980).

\bibitem{effaction} L.~Parker and D.~Toms, {\sl Quantum field theory in curved spacetime}\/(Cambridge Univ. Press, Cambridge, 2009).

\bibitem{tpaspects}
T.~Padmanabhan, {\sl Aspects of Quantum Field Theory} in `Geometry, Fields and Cosmology', ed. Iyer and Vishweshwara (Kluwer Academic Publishers, 1997).

\bibitem{pexpuniv} 
L.~Parker, Phys.\ Rev.\ Lett.,~ {\bf 21}, 562 (1968);
N.~D.~Birrell and P.~C.~W.~Davies, {\sl Quantum Fields in Curved Space}\/ (Cambridge Univ. Press, Cambridge, 1982).

\bibitem{bhevap}
S.~W.~Hawking, Nature,~{\bf 248}, 30 (1974);
S.~W.~Hawking, Commun.\ Math.\ Phys.,~{\bf 43}, 199 (1975);
J.~B.~Hartle and S.~W.~Hawking, Phys.\ Rev.,~{\bf D13}, 2188 (1976);
T.~Padmanabhan, Phys.\ Rept.,~{\bf 406}, 49 (2005), [arXiv:gr-qc/0311036v2].


\bibitem{euhe}
W.~Heisenberg and H.~Euler, Zeitschr, Phys.,~{\bf98}, 714 (1936) translated by W.~Korolevski and H.~Kleinert, [arXiv:physics/0605038v1];
W.~Dittrich, M.~Reuter, {\sl Effective Lagrangians in Quantum Electrodynamics}, Lect.\ Notes\ Phys. 220 (1985); 
S.~Blau, M.~Visser and A.~Wipf, Int.\ J.\ Mod.\ Phys.,~{\bf A6}, 5409 (1991), [arXiv:hep-th/0906.2851v1];
T.~Padmanabhan, Pramana,~\textbf{37}, 179 (1991);
R.~Soldati and L.~Sorbo, Phys.\ Lett.,~{\bf B426}, 82 (1998), [arXiv:hep-th/9802167v1];
G.~V.~Dunne and C.~Schubert, Nucl.\ Phys.,~{\bf B564}, 591 (2000), [arXiv:hep-th/9907190v1];
U.~D.~Jentschura, H.~Gies, S.~R.~Valluri, D.~R.~Lamm and E.~J.~Weniger, Can.\ J.\ Phys.~{\bf 80}, 267 (2002), [arXiv:hep-th/0107135v2]; 
G.~V.~Dunne, in {\sl`From Fields to Strings: Circumnavigating Theoretical Physics'}, ed. M.~Shifman et al., (World Scientific, 2005), [arXiv:hep-th/0406216v1].

\bibitem{brown}
M.~R. Brown, {\sl Quantum Gravity at Small Distances} in `Quantum Theory of Gravity',  ed.  M.~S.~Christensen, (Adam Hilger Ltd., 1994).

\bibitem{Alexanian} 
G.~Alexanian, R.~MacKenzie, M.~B.~Paranjape and J.~Ruel, Phys.\ Rev.~{\bf D77}, 105014 (2008), [arXiv:hep-th/0802.0354v2].

\bibitem{ll1}
L.~D.~Landau and E.~M.~Lifshitz, {\sl Mechanics}, Volume 1 of {\sl Course of Theoretical Physics}, Elsevier.

\bibitem{TPSreeni}
K.~Srinivasan and  T.~Padmanabhan, Phys.\ Rev.~{\bf D60}, 024007 (1999), [arXiv:gr-qc/9812028v1].

\bibitem{factortwo}
S.~Shankaranarayanan, K.~Srinivasan and T.~Padmanabhanan, Mod.\ Phys.\ Letts.,~{\bf 16},  571 (2001), [arXiv:gr-qc/0007022v2];
S.~Shankaranarayanan, T.~Padmanabhan and K. ~Srinivasan, Class.\ Quan.\ Grav,~{\bf 19}, 2671 (2002), [arXiv:gr-qc/0010042v4];
E.~C.~Vagenas, Nuovo\ Cim.,~{\bf B117}, 899 (2002), [arXiv:hep-th/0111047v3];
S.~Shankaranarayanan, Phys.\ Rev.,~{\bf D67}, 084026 (2003), [arXiv:gr-qc/0301090v2];
T.~Padmanabhan, Mod.\ Phys.\ Lett.,~{\bf A19}, 2637 (2004), [arXiv:gr-qc/0405072v2];
E.~T.~Akhmedov, V.~Akhmedova and D.~Singleton, Phys.\ Lett.,~{\bf  B642}, 124 (2006), [arXiv:hep-th/0608098v2];
E.~T.~Akhmedov, V.~Akhmedova, T.~Pilling and D.~Singleton, Int.\ J.\ Mod.\ Phys,~{\bf  A22}, 1705 (2007), [arXiv:hep-th/0605137v4];
P.~Mitra, Phys.\ Lett.,~{\bf B648}, 240 (2007), [arXiv:hep-th/0611265v3];
S.~P.~Kim, JHEP, 0711:048 (2007), [arXiv:hep-th/0710.0915v2];
S.~P.~Kim, J.\ Korean\ Phys.\ Soc.,~{\bf 53}, 1095 (2008), [arXiv:hep-th/0709.4313v1];
T.~Pilling, Phys.\ Lett.,~{\bf B660}, 402 (2008), [arXiv:gr-qc/0709.1624v4];
R.~Banerjee and B.~R.~Majhi, JHEP, 0806:095 (2008), [arXiv:hep-th/0805.2220v2 ];
B.~Chatterjee, A.~Ghosh and P.~Mitra, Phys.\ Lett.,~{\bf B661}, 307 (2008), [arXiv:hep-th/0704.1746v4];
C.~Ding, M.~Wang and J.~Jing, Phys.\ Lett.,~{\bf B676}, 99 (2009);
Criscienzo et al., [arXiv:gr-qc/0906.1725v2], (2009);
X.~Liu and W.~Liu, Int.\ J.\ Theor.\ Phys.,~{\bf 48}, 3614 (2009); 
C.~Ding and J.~Jing, Class.\ Quantum\ Grav.,~{\bf 27}, 035004 (2010), [arXiv:gr-qc/1001.2946v2];
Y.~Chen and K.~Shao, [arXiv:hep-th/1007.4367v2], (2010);
Y.~Hu, J.~Zhang and Z.~Zhao, Mod.\ Phys.\ Lett.,~{\bf A25}, 295 (2010);
A.~Yale, [arXiv:gr-qc/1012.2114v2], (2010);
J.~T.~Firoujaee and R.~Mansouri, [arXiv:grqc/1104.0530v1], (2011).

\bibitem{mukhanov}
V.~Mukhanov and S.~Winitzki, {\sl Introduction to Quantum Effects in Gravity}, (Cambridge Univ. Press, Cambridge, 2007).

\bibitem{grad}
I.~S.~Gradshteyn and I.~M.~Ryzhik, {\sl Table of Integrals, Series and Products}\/ (Academic Press, London, 1994).

\bibitem{tunnelingreview} L. Vanzo, G. Acquaviva, R. Di Criscienzo, [arXiv:gr-qc/1106.4153v1], (2011).

\bibitem{rmsphoton} 
L.~D.~Landau and E.~M.~Lifshitz, {\sl Mechanics}, Volume 5 of {\sl Course of Theoretical Physics}, p.346 (Elsevier);
T.~Padmanabhan, {\sl Theoretical Astrophysics - Astrophysical Processes}, Volume 1, p.215 (Cambridge Univ. Press, Cambridge, 2000).

\bibitem{tpplanewavesacc}
K.~Srinivasan, L.~Sriramkumar and  T.~Padmanabhan,  Phys.\ Rev. {\bf D56}, 6692 (1997).

\bibitem{trick}
D.~Khandekar, S.~Lawande and K.~Bhagwat, {\sl Path-integral methods and their applications}\/ (Allied Publishers, 2002).

\end{thebibliography}
\end{document}